\documentclass[twocolumn]{aastex63}
\usepackage{graphicx,bm,amssymb,amsmath,times}
\usepackage{xfrac}
\numberwithin{equation}{section}
\usepackage{xcolor}
\usepackage{gensymb}

\usepackage{color}
\usepackage{acronym}   
\definecolor{ochre}{rgb}{0.8, 0.47, 0.13}
\definecolor{cyan}{rgb}{0, 0.7, 1}

\acrodef{BH}{black hole}
\acrodef{HMXB}{high-mass X-ray binary}
\acrodefplural{HMXB}[HMXBs]{high-mass X-ray binaries}

\acrodef{RLOF}{Roche-lobe overflow}
\acrodef{CE}{common-envelope}

\acrodef{MS}{main sequence}
\acrodef{HeMS}{helium main sequence}

\acrodef{MT}{mass transfer}

\acrodef{WR}{Wolf-Rayet}

\received{\today}

\shorttitle{WR winds and Cygnus X-1}
\shortauthors{Neijssel et al.}

\begin{document}
\title{Wind mass-loss rates of stripped stars inferred from Cygnus X-1}

\author{Coenraad J. Neijssel} \email{cneijssel@star.sr.bham.ac.uk}
\affiliation{Birmingham Institute for Gravitational Wave Astronomy and School of Physics and Astronomy,\\
University of Birmingham, Birmingham, B15 2TT, United Kingdom}
\affiliation{Monash Centre for Astrophysics, School of Physics and Astronomy, Monash University, Clayton, Victoria 3800, Australia}
\affiliation{The ARC Center of Excellence for Gravitational Wave Discovery -- OzGrav}
\author{Serena Vinciguerra}
\affiliation{Max Planck Institute for Gravitational Physics (Albert Einstein Institute), D-30167 Hannover, Germany}
\affiliation{Monash Centre for Astrophysics, School of Physics and Astronomy, Monash University, Clayton, Victoria 3800, Australia}
\affiliation{The ARC Center of Excellence for Gravitational Wave Discovery -- OzGrav}
\affiliation{Anton Pannekoek Institute for Astronomy, University of Amsterdam, Science Park 904, 1090GE Amsterdam, The Netherlands}
\author{Alejandro Vigna-G\'{o}mez}
\affiliation{DARK, Niels Bohr Institute, University of Copenhagen, Blegdamsvej 17, 2100 Copenhagen, Denmark}
\author{Ryosuke Hirai}
\affiliation{Monash Centre for Astrophysics, School of Physics and Astronomy, Monash University, Clayton, Victoria 3800, Australia}
\affiliation{The ARC Center of Excellence for Gravitational Wave Discovery -- OzGrav}
\author{James C.~A.~Miller-Jones}
\affiliation{International Centre for Radio Astronomy Research -- Curtin University, GPO Box U1987,
Perth, WA 6845, Australia}
\author{Arash Bahramian}
\affiliation{International Centre for Radio Astronomy Research -- Curtin University, GPO Box U1987,
Perth, WA 6845, Australia}
\author{Thomas J.~Maccarone}
\affiliation{Department of Physics \& Astronomy, Texas Tech University, Box 41051, Lubbock TX, 79409-1051, USA}
\author{Ilya Mandel} \email{ilya.mandel@monash.edu}
\affiliation{Monash Centre for Astrophysics, School of Physics and Astronomy, Monash University, Clayton, Victoria 3800, Australia}
\affiliation{The ARC Center of Excellence for Gravitational Wave Discovery -- OzGrav}
\affiliation{Birmingham Institute for Gravitational Wave Astronomy and School of Physics and Astronomy,\\
University of Birmingham, Birmingham, B15 2TT, United Kingdom}

\begin{abstract}
Recent observations of the high-mass X-ray binary Cygnus X-1 have shown that both the companion star (41 solar masses) and the black hole (21 solar masses) are more massive than previously estimated. Furthermore, the black hole appears to be nearly maximally spinning. Here we present a possible formation channel for the Cygnus X-1 system that matches the observed system properties. In this formation channel, we find that the orbital parameters of Cygnus X-1, combined with the observed metallicity of the companion, imply a significant reduction in mass loss through winds relative to commonly used prescriptions for stripped stars.  
\end{abstract}



\section{Introduction}

Cygnus X-1 is a \ac{HMXB} in the Cygnus OB3 association, which hosts a star in orbit with a \ac{BH} \citep[e.g.][]{Webster:1972,Bolton:1972,Bolton:1975,Hutchings:1973}. The \ac{BH} accretes matter from the stellar wind; this accretion powers X-ray radiation  \citep[e.g.][]{Davidson:1973,vdHeuvel:1975,Conti:1978,Petterson:1978} and a jet \citep[e.g.][]{Bisiacchi:1974,Marti:1996,Stirling:2001}. \citet{Orosz:2011} inferred the \ac{BH} and stellar companion masses of Cygnus X-1 to be $14.8 \pm 1.0 M_\odot$ and $19.2 \pm 1.9 M_\odot$, respectively.  Revised measurements of the distance to Cygnus X-1 \citep{MillerJones:2020} indicate that both objects are significantly more massive. The temperature and luminosity of the optical companion are estimated to be $T_\mathrm{eff}=31.1\pm0.7$ kK and $\mathrm{log}(L/L_{\odot}) = 5.63\pm 0.07$ with a mass of  $M_\mathrm{opt}=40.6_{-7.1}^{+7.7}M_{\odot}$, where we quote the median value and the 68 per cent confidence interval boundaries \citep{MillerJones:2020}. The mass of the \ac{BH} is estimated as $M_\mathrm{BH}=21.2_{-2.3}^{+2.2}M_{\odot}$. The binary has an almost circular orbit with a semi-major axis $a=0.244_{-0.013}^{+0.012}$ AU and eccentricity $e=0.0189_{-0.0026}^{+0.0028}$ \citep{MillerJones:2020}. The \ac{BH} is inferred to be nearly  maximally spinning with a dimensionless spin of at least 0.95 according to both disk continuum and reflection line fitting studies \citep{Gou:2011,Fabian:2012,MillerJones:2020,Zhao:2020}. Via optical spectroscopy, it has been found that the ratios of the surface abundances of both helium and iron to hydrogen are about twice the respective values for the Sun \citep{Shimanskii:2012}.  As we discuss below, these observations taken together present a challenge for models of massive stellar binary evolution.

In this paper we describe the constraints that these observations place on the Cygnus X-1 formation channel (Sec.~\ref{sec:channel}). We explore how the \ac{HeMS} phase of this channel provides a constraint on the wind mass-loss rates of massive stars (Sec.~\ref{sec:winds}). We present the likely future of the system and its implications for gravitational-wave detections (Sec.~\ref{sec:future}). Finally we discuss some caveats and questions raised by the observations of Cygnus X-1 (Sec.~\ref{sec:discussion}).

\section{Cygnus X-1: Observations and Assumed Channel}\label{sec:channel}
\begin{figure}
\centering
\includegraphics[width=\columnwidth]{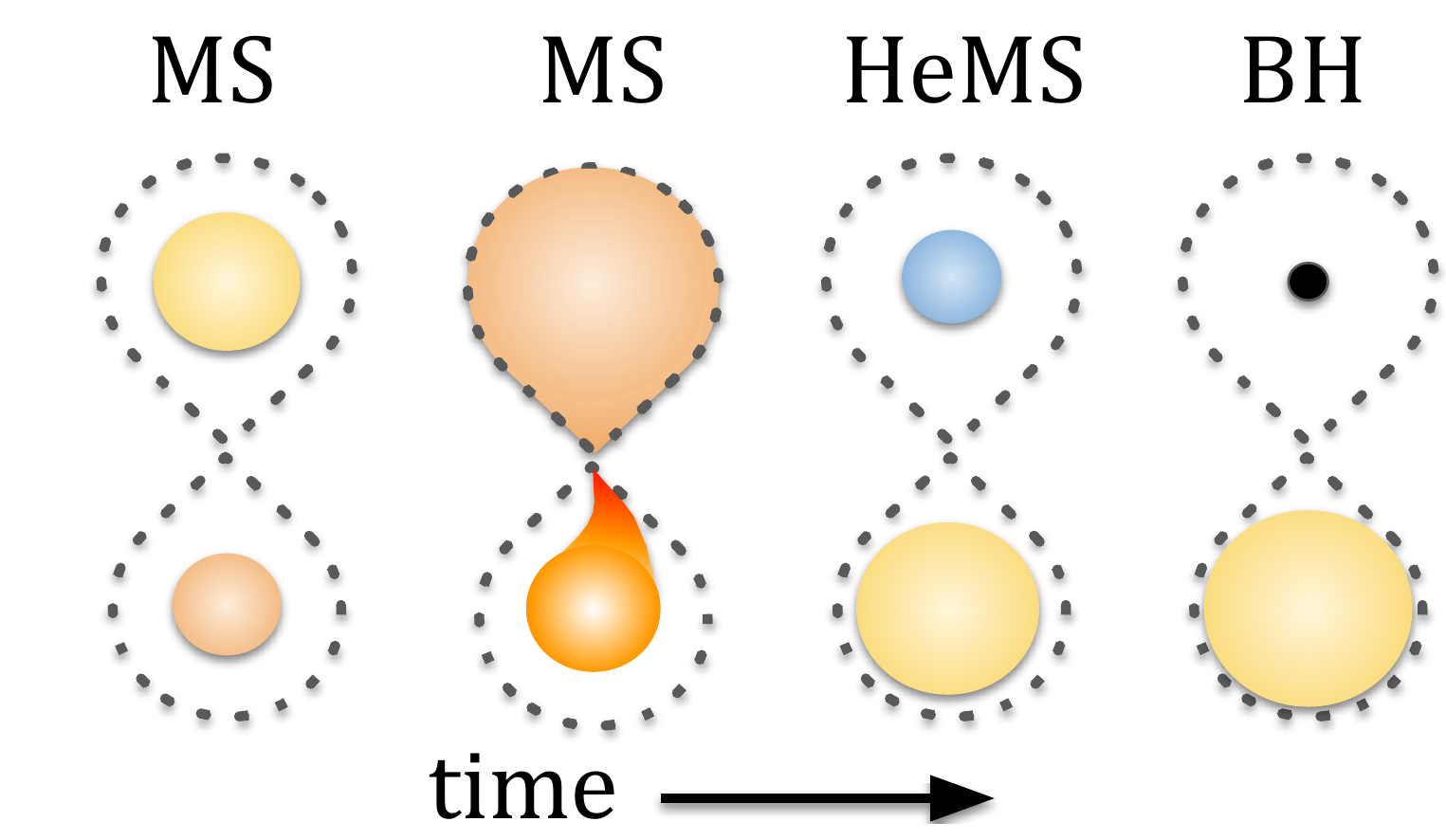}
   \caption{Assumed formation channel for Cygnus X-1. 1: Two stars (primary top, secondary bottom) are born in a binary. 2: The primary star is more massive and evolves faster, expands and starts stable mass transfer onto the secondary. 3: The primary is left as a hot stripped \ac{HeMS} star with a companion which could have accreted a significant amount. 4: The primary star collapses and leaves behind a \ac{BH} orbiting the secondary \ac{MS} star.}
    \label{fig:prior}
\end{figure}

We assume the following formation channel for Cygnus X-1 (see Fig.\ref{fig:prior} for illustration). Two stars are born in a binary. The more massive star (the primary) evolves more quickly and expands first. The companion (the secondary) is close enough for the late main sequence primary to commence mass transfer. The primary is stripped of its envelope, leaving an exposed helium core. The core continues nuclear fusion until it collapses and forms a \ac{BH} in orbit with the still core-hydrogen-burning \ac{MS} companion. In the sections below, we describe the observations and theoretical analyses that support this channel.

\subsection{Eccentricity, peculiar velocity and fallback}\label{sec:SN}

The collapse of a star into a compact object can impart both a natal kick from an asymmetric explosion (see e.g. \citealt{Fryer:2004} and references therein) and a kick from rapid symmetric mass loss to the system \citep{Blaauw:1961}.  The low eccentricity of the Cygnus X-1 binary \citep{Orosz:2011,MillerJones:2020} seems to disfavour a significant natal kick.  However, tidal forces could have circularised the system since the collapse: for the inferred stellar and binary properties, the circularisation timescale for dynamical tides \citep[applicable for radiative-envelope stars,][]{Zahn:1977} is only $\sim 10^5$ years \citep[estimated using equations (41) and (43) of][]{Hurley:2002}.

On the other hand, the system's small peculiar velocity strongly indicates that the \ac{BH} experienced nearly complete collapse with little mass ejection during its formation \citep{Mirabel:2003}.  The peculiar velocity is $10.7\pm2.7$ km s$^{-1}$ relative to its host association \citep{Rao:2020}, which limits the amount of instantaneous symmetric mass loss \citep{Blaauw:1961, Nelemans:1999, Wong:2012} to $\lesssim 2 M_{\odot}$, unless the kick from symmetric mass ejection is fortuitously cancelled by an oppositely directed natal kick.  Further indirect evidence for the low natal kick lies in the absence of Type C quasi periodic oscillations, which may indicate Lense-Thirring precession due to spin-orbital misalignment as a result of a natal kick \citep{StellaVietri:1998} and are observed in most known black hole X-ray binaries \citep{Ingram:2016}, the majority of which are believed to have received strong ($\sim 100\mathrm{\,\, km\, s^{-1}}$) natal kicks \citep{Atri:2019}.  The low amount of ejected mass is also consistent with the current eccentricity in the absence of significant tidal circularisation, and suggests nearly complete fallback onto the black hole except for a small amount of neutrino mass loss \citep[e.g.][]{Nadezhin:1980,Lovegrove:2013,Fernandez:2018}.  Nearly complete collapse matches the theoretical models simulating the fall-back onto black holes of similar masses \citep{Fryer:2012} and observational evidence for massive stars disappearing without supernovae \citep{Adams:2017}.  

\subsection{Black-hole progenitor mass}
The Eddington limit on the mass accretion rate for a \ac{BH} of this mass is $\lesssim  2 \times 10^{-7} M_{\odot}$ yr$^{-1}$.  Under the assumption of Eddington-limited accretion (but see, e.g., \citealt{FragosMcClintock:2015,Eldridge:2017,vanSon:2020}, who relax this assumption),
only a negligible amount of mass could have been accreted onto the \ac{BH} since it formed \citep{KingKolb:1999}.  The companion lifetime sets the accretion duration to no more than a few Myr, meaning that at most $\lesssim \mathrm{few} \times 10^{-1}M_{\odot}$ could have been accreted.  Assuming that the jet turned on promptly after the formation of the black hole, the few$\times 10^4$ yr estimate of the age of the jet \citep{Russell:2007} places an even stronger constraint on the amount of accreted mass, $\lesssim 10^{-2} M_{\odot}$ (see also \citealt{Sell:2015}, who consider other models for the nebula origin but reach similar conclusions about its age).  Therefore, the current black hole mass $M_\mathrm{BH}=21.2_{-2.3}^{+2.2}M_{\odot}$ is a good estimate for the progenitor mass just before the collapse.

\subsection{Black hole spin \& progenitor spin}
The \ac{BH} has a dimensionless spin $\chi$ close to unity, $\chi \geq 0.95$ \citep{Gou:2011,Fabian:2012,MillerJones:2020}, although spin measurements may have large systematic errors because of  modelling assumptions \citep{MillerMiller:2015,Kawano:2017,Zhao:2020}.   A spin close to unity, coupled with negligible mass loss during collapse, suggests that if the pre-collapse progenitor had excess angular momentum, $\chi>1$, it must have been carried away by a small amount of mass with high specific angular momentum \citep[cf.][]{Janka:2013,Janka:2017,Batta:2019,Murguia:2020}. 

The rotational angular momentum of the \ac{BH} was either present in the progenitor, or was gained during or after its collapse. Here we briefly summarise why we assume that the angular momentum comes from the progenitor and why this implies that the progenitor must have been stripped early in its evolution (see \citealt{MandelFragos:2020} for a longer discussion).

A \ac{BH} needs to roughly double its mass through accretion in order to go from zero to maximal spin \citep{Bardeen:1973, Thorne:1974}. This is not possible for Cygnus X-1 under the assumption of Eddington-limited accretion.  The spin could have been acquired during the collapse, e.g., if the companion torques some of the ejecta which then fall back onto the \ac{BH} \citep{Batta:2017,Schroder:2018}, but this requires some fine-tuning: for example, the ejecta must have sufficient velocities to be torqued by the companion, but without a significant tail of escaping ejecta to match formation through nearly complete fallback as discussed above.  Therefore we assume that the angular momentum was present in the progenitor at the moment of the collapse.

Observations suggest that stars might be born as rapid rotators \citep{Fukuda:1982, Rosen:2012,Ramirez-Agudelo:2013, Ramirez-Agudelo:2015}, although these observations could be affected by binary interactions \citep{Langer:2008, deMink:2013}. Even if stars are born as rapid rotators, they spin down through wind-driven mass loss and are unlikely to retain enough angular momentum to form rapidly spinning \acp{BH}. 

The black hole's dimensionless spin is determined by the ratio of its total angular momentum to the square of its mass.  At lowest order, it is therefore insensitive to the redistribution of angular momentum through the progenitor star.  However, if the bulk of the angular momentum moves into the envelope as the star expands, and the envelope is subsequently removed by winds or mass transfer, the remaining core is likely left with too little angular momentum to produce a rapidly spinning \ac{BH} \citep{Petrovic:2005,Belczynski:2017,FullerMa:2019,Bavera:2019,MandelFragos:2020}.  On the other hand, the core could retain sufficient angular momentum to form a rapidly spinning \ac{BH} if there is a large amount of differential rotation between the layers of the star  \citep{Hirschi:2005}.  Therefore, the efficiency of angular momentum transport in the star plays a key role, yet both the mechanism and degree of core-envelope coupling remain uncertain.  Theory \citep{Tayler:1973,Spruit:2002,Fuller:2019,Takahashi:2020} and observations, such as rotation rates of low-mass giant stars \citep{Cantiello:2014}, suggest that there may be efficient angular momentum transport and the envelope is coupled to the core.  If so, the observed rapid spin of the Cygnus X-1 black hole in a close binary, in which the progenitor must have lost its envelope, appears to require pre-collapse interaction with the companion to spin up the stellar core.

Tidal locking of the period of the stellar rotation to the period of the binary provides the most likely mechanism for producing a rapidly rotating \ac{BH} progenitor \citep{Izzard:2004tides, Kushnir:2016,Zaldarriaga:2017,Belczynski:2017,Bavera:2019}.  Chemically homogeneous evolution could yield rapidly rotating black holes \citep{MandelDeMink:2016,Marchant:2016}, but is not expected to operate at such high metallicities and is not consistent with the observed expansion of the companion.  

Instead, the following evolutionary sequence, proposed by \citet{Valsecchi:2010} for M33 X-7, a similar high-mass X-ray binary with a rapidly spinning \ac{BH}, and investigated in detail by \citet{Qin:2019}, appears to be the most likely formation mechanism for Cygnus X-1.  The binary starts out with a period somewhat shorter than the current observed one.  The more massive primary -- the \ac{BH} progenitor -- commences mass transfer while still on the main sequence.  This mass transfer, is likely to be largely non-conservative if limited by the spin-up of the accretor \citep{Qin:2019}; this is favoured both by the observed evolutionary state of the secondary (conservative mass transfer would imply an initially lower-mass primary, and hence an older binary, placing an upper limit on the \ac{BH} progenitor mass that would make it unlikely to form such a massive \ac{BH}) and the observed period.  The mass transfer removes the donor's envelope, preventing subsequent re-expansion and angular momentum loss.  Meanwhile, the core remains tidally locked on the \ac{MS}.  After hydrogen is exhausted in the core at the end of the \ac{MS}, the star contracts into a rapidly spinning \ac{HeMS} star.  While this \ac{HeMS} star is no longer tidally locked, it can still collapse into a rapidly spinning \ac{BH}.  \citet{Qin:2019} find that the efficiency of angular momentum transport does not play a significant role for the evolution of the black hole progenitor during the main sequence, where tidal locking keeps its core rapidly spinning, but could be a key factor in determining the ultimate black hole spin through its impact on the angular momentum lost through winds in later evolutionary stages.  The \citet{Qin:2019} models successfully reproduce systems with the orbital parameters of Cygnus X-1.  In fact, the match appreciably improves with the upward revision in the \ac{BH} and optical companion masses, as the observed masses in the bottom left panel of figure 3 of \citet{Qin:2019} shift toward the locus of their model evolutionary trajectories (though their evolutionary trajectories are only shown for an initial mass ratio of $0.4$, while the latest observations support more comparable masses).  Given the number of uncertainties relating to this first mass transfer episode, in what follows we focus on the subsequent evolution of the binary under the assumption of this formation channel for Cygnus X-1.

\subsection{Companion mass \& abundance}\label{sec:companion}

We compared the observed mass, luminosity and temperature of the optical companion against analytic fits to stellar tracks of \citet{Hurley:2000} as implemented in the COMPAS rapid population synthesis code \citep{Stevenson:2017,VignaGomez:2018}.  The observations are consistent with a \ac{MS} star that is about 70 to 80 per cent through its core hydrogen burning phase, ignoring the impact of rotation. 

These stellar tracks are for stars with regular hydrogen-rich atmospheres.  Using them ignores the possible impact of non-standard surface abundances, and so corresponds to the assumption that only a thin surface layer has a significant over-abundance of helium, rather than a uniform distribution of enriched material throughout the star.  Stars in later stages of the \ac{MS} with the relevant mass and metallicity should have at most a very thin convective layer at the surface \citep[e.g.,][]{Maeder:2008,Kippenhahn:2012}.  Therefore, mixing is expected to be relatively inefficient: in the absence of large-scale convection, the Rayleigh-Taylor instability is likely to be suppressed by temperature inversion in the accreted material \citep{Kippenhahn:1980,BraunLanger:1995}. The resulting thermohaline mixing will mix the helium-rich material through the companion only on timescales longer than the expected few$\times 10^4$ years since the formation of the \ac{BH} if the bulk of the enriched material was accreted at or shortly before the \ac{BH} formation.  

The evolutionary channel shown in figure \ref{fig:prior} assumes that the optical companion has not overflowed its Roche lobe.  While previous studies explored this possibility, perhaps with intermittent Roche lobe overflow followed by longer periods when the binary is detached as in the present state, these models were typically based on assumptions that the companion is less massive than the \ac{BH} \citep{Podsiadlowski:2003}, which is inconsistent with present observations.  Indeed, if we assume that mass transfer onto a black hole is almost entirely non-conservative because of the Eddington limit, the binary's semi-major axis $a$ evolves as
\begin{equation}\label{eq:a_MT}
    \frac{\dot{a}}{a} = -2 \frac{\dot{M}_\mathrm{opt}}{M_\mathrm{opt}}\left[1-\left(\gamma+\frac{1}{2}\right)\frac{M_\mathrm{opt}}{M_\mathrm{opt}+M_\mathrm{BH}}\right],
\end{equation}
where $\gamma$ is the specific angular momentum of the ejected material in units of the binary's specific orbital angular momentum $J/(M_\mathrm{opt}+M_\mathrm{BH})$.  The binary can widen as a result of such mass transfer only if $\gamma \lesssim 1$ for the observed component masses.  However, in the common assumption of isotropic re-emission from the accretor, $\gamma = M_\mathrm{opt}/M_\mathrm{BH} \approx 2$.  The ejected material would have to carry much lower specific angular momentum in order for the binary to be able to detach once Roche lobe overflow from the companion commences, which seems unlikely, lending support to our proposed channel.

The estimated mass and lifetime of the companion already enable us to roughly infer the initial mass of the \ac{BH} progenitor.  We assume that both stars are born and start fusion at the same time and use a fit \citep{Farr:2018} to the \citet{Brott:2011,Kohler:2015} stellar models for the \ac{MS} lifetime of non-rotating massive stars.  For this simple estimate, we ignore the impact of binary interactions, consistent with the assumption of largely non-conservative mass transfer.
The \ac{BH} progenitor should have had an initial mass of $\sim 55$ -- $75 M_\odot$ in order to complete its evolution while leaving behind a \ac{MS} companion of mass $M_\mathrm{opt}$ at $\sim 70\%$ -- $80\%$ of its \ac{MS} lifetime. 

The surface abundances of the companion are non-standard for massive \ac{MS} stars and a challenge to explain even in the context of binary interaction.  We discuss the abundances in more detail in Section \ref{sec:discussion}.  Here, we focus on the iron abundance, as this is key to the analysis of line-driven winds in the following section.  \citet{Shimanskii:2012} find that the iron abundance of the companion is 2.2 times the solar iron abundance, although precise measurements are challenging due to the complexity of the system.  Assuming $Z_\odot = 0.014$ for solar metallicity \citep{Asplund:2009}, this corresponds to an effective metallicity of $Z \approx 0.03$.  On the other hand, \citet{Daflon:2001} find a slightly sub-solar iron abundance of $\log \epsilon(\mathrm{Fe}) = 7.33\pm0.12$ in HD~227460, which is a B0.5V star in the same Cygnus OB3 association as Cygnus X-1.  This could indicate that $Z \approx 0.01$ is a better estimate of the initial iron abundance, and the iron abundance of the companion in Cygnus X-1 has been enhanced during the collapse of the primary to a \ac{BH}.  Therefore, we explore an initial metallicity range $0.01 \leq Z \leq 0.03$ in the following section.

\section{Cygnus X-1: Maximum wind mass-loss rate}\label{sec:winds}

Hereafter we assume that, after the mass transfer episode induced by the \ac{BH} progenitor,  the primary is left with no hydrogen layer on top of the He core. This assumption is consistent with the channel proposed by \citet{Qin:2019}, in which the rapidly spinning \ac{BH} is ultimately formed through the collapse of a Wolf-Rayet star (see however discussion in Section \ref{sec:discussion}).

\begin{figure}
\centering
\includegraphics[width=\columnwidth]{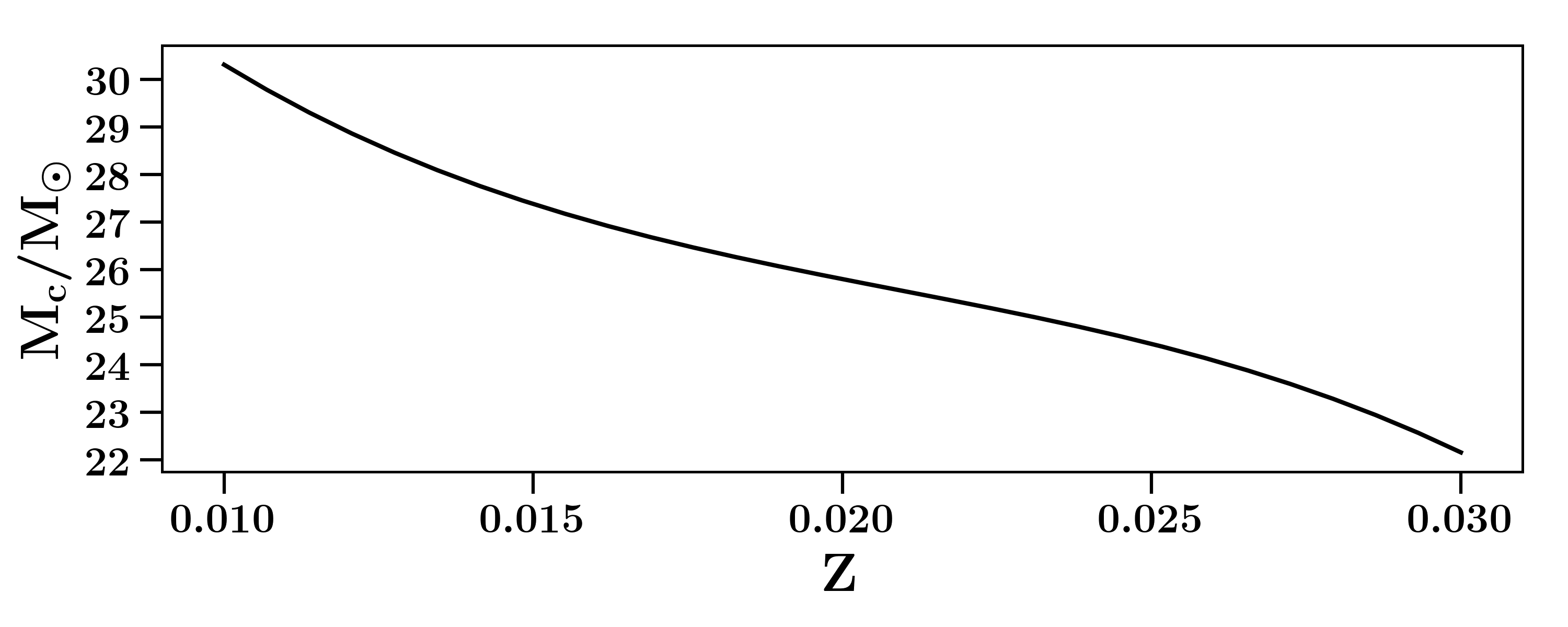}
   \caption{
   Maximum helium core mass at terminal-age \ac{MS} as a function of metallicity, maximized over zero-age \ac{MS} masses, based on the stellar tracks of \citet{Hurley:2000} as implemented in COMPAS \citep{Stevenson:2017,VignaGomez:2018}.
   }
    \label{fig:TAMS_cores}
\end{figure}

In order for Cygnus X-1 to form through the channel depicted in Fig.~\ref{fig:prior}, two things must be  true.  Firstly, the \ac{HeMS} star must be born with sufficient mass to give rise to the observed \ac{BH} mass even after losing mass through Wolf-Rayet winds during the \ac{HeMS}. However, at a given metallicity, there is a maximum to the He core mass that can be formed at the end of the \ac{MS}: as more massive stars have higher wind mass loss rates, terminal-age \ac{MS} He core masses asymptotically approach a maximum as a function of zero-age \ac{MS}.  This maximum He core mass is plotted in figure \ref{fig:TAMS_cores}.  Because the \ac{BH} progenitor is stripped of its hydrogen envelope at the end of the \ac{MS} for the binary evolutionary channel we assume, we can directly constrain the amount of mass loss during the \ac{HeMS} phase.  The Wolf-Rayet winds must then not remove more than the difference between this maximum mass and the final \ac{BH} mass.  We refer to this as the  {\it ``$M_\mathrm{HeMS} \leq M_\mathrm{HeMS,max}$''} condition.

The maximum He core mass plotted in figure \ref{fig:TAMS_cores} is very sensitive to the \ac{MS} wind prescriptions (see discussion in \citealt{Renzo:2017,Neijssel:2019,MillerJones:2020}).  The analysis of the remnant mass from massive single stars is particularly sensitive to luminous blue variable winds \citep{Belczynski:2010,MillerJones:2020}, which can remove the hydrogen envelope of the star and hasten the onset of the Wolf-Rayet phase \citep{Conti:1975}; therefore, increasing the potential remnant mass can be achieved by decreasing either luminous blue variable winds or Wolf-Rayet winds. The plot in figure \ref{fig:TAMS_cores} assumes the default COMPAS luminous blue variable mass loss rate of $1.5 \times 10^{-4} \mathrm{M}_\odot$ yr$^{-1}$ \citep{Belczynski:2010} for stars approaching the Humphreys-Davidson limit \citep{HumphreysDavidson:1994}. The convective overshooting parameter employed in stellar evolution calculations is another source of uncertainty, with greater overshooting leading to larger cores \citep{Brott:2011}.  To explore the impact of convective overshooting, we simulated the He core mass at terminal-age \ac{MS} for stars with different metallicities starting from a zero-age \ac{MS} mass of 150 M$_\odot$ using the stellar evolution code MESA (v12115, \citealt{Paxton:2011}). We find that varying the overshooting parameter by an order of magnitude between $f=0.02$ and $f=0.2$ (\citealt{Qin:2019} used $f=0.11$) with the step overshoot scheme changes the He core mass at terminal-age \ac{MS} by between 1\% and 13\% for the range of metallicities in figure \ref{fig:TAMS_cores}.  Finally, the COMPAS single stellar evolution models are based on the fitting formulae of \citet{Hurley:2000} to the evolutionary tracks of \citet{Pols:1998} which involve extrapolations to higher masses than those covered by the initial range of models, and may under-estimate the He core masses of massive stars.  We therefore use the following additional constraint, which bypasses these sources of uncertainty in the maximum He core mass.

In our assumed channel the secondary \ac{MS} companion must not overflow its Roche lobe onto the \ac{HeMS} primary.  We write this constraint as \citep{Eggleton:1983}
\begin{equation}\label{eq:RocheRadiusCondition}
R_2(t) \leq a(t) \frac{0.49 q^{2/3}(t)}{0.6q^{2/3}(t) + \ln(1+q^{1/3}(t))} , 
\end{equation}
where $m_{\rm 2}(t)$ and $R_{\rm 2}(t)$ are the mass and radius of the companion as a function of time, $m_{\rm 1}(t)$ the mass of the \ac{BH} or its progenitor, $q(t)=\frac{m_2(t)}{m_1(t)}$, is the mass ratio, and $a(t)$ the orbital separation of the binary.  

The orbital separation widens due to wind mass loss.  In the limit of fast, non-interacting winds, the widening is described by
\begin{equation}\label{eq:wideningWinds}
    \frac{\dot{a}}{a} = - \frac{\dot{M_\mathrm{tot}}}{M_\mathrm{tot}},
\end{equation}
where $M_{\rm tot} = m_{\rm 1}+m_{\rm 2}$ is the total mass of the system.  Because winds widen the binary and remove mass from the \ac{HeMS} primary faster than from its \ac{MS} companion, they increase the size of the secondary's Roche lobe over time.  Thus, even though the secondary is not overflowing  its Roche lobe now, it may have done so in the past if the mass-loss rate was high.  If we evolve the system back in time, the requirement that the secondary never overflows its Roche lobe imposes an alternative upper limit \citep{Axelsson:2011} on the maximum Wolf-Rayet wind mass-loss rate.  We refer to this as the {\it ``no \ac{RLOF}''} condition. 
Although the non-interacting wind assumption, describing the widening of the binary (Equation \ref{eq:wideningWinds}), may be an over-simplification for such short-period systems \citep{MacLeodLoeb:2020}, it allows us to conservatively estimate the constraints imposed by the existence of Cygnus X-1 and presented below.

We parametrise the mass-loss rate through Wolf-Rayet winds, modelled with the prescription proposed by \citet{Belczynski:2010} and based on \citep{Hamann:1995,Hamann:1998,Belczynski:2010}, with a multiplicative parameter $f_\mathrm{WR}$, following \citet{Barrett:2017FIM} (see Appendix \ref{app:winds} for a definition and discussion). 
We constrain the allowed parameter space of the wind strength by rewinding the evolution of the binary from the current state. Because the \ac{BH} formed recently in our model, we set the luminosity and temperature of the \ac{MS} companion at the end of the \ac{HeMS} phase of the primary equal to the current inferred luminosity and temperature of the observed \ac{MS} secondary. The mass and age of the secondary inferred from temperature and luminosity vary slightly for different metallicities. As we argued earlier, the \ac{BH} is expected to lose negligible mass during collapse, so we set the mass of the HeMS primary at the end of that phase equal to the inferred \ac{BH} mass.  We assume that the secondary was 99.7\% Roche-lobe filling at the end of the primary's HeMS phase \citep{MillerJones:2020}\footnote{In \citet{MillerJones:2020} the Roche-lobe filling factor is defined as the ratio between the distance from the center of the star to the point where the equipotential surface enclosing the volume of the star crosses the axis connecting the star to its \ac{BH} companion, and the distance from the center of the star to the L1 Lagrange point.  In this paper, we define the Roche-lobe filling factor as the ratio between the radius of the star and the volume-equivalent Roche-Lobe radius of  \citet{Eggleton:1983}. We therefore convert the median and one-$\sigma$ lower bounds of 0.96 and 0.93 reported in table 1 of \citet{MillerJones:2020} to values of 0.997 and 0.99, respectively, that we use here.}. 

The reverse evolution of the \ac{MS} and \ac{HeMS} stars is followed using the analytic fits to the stellar tracks of \citet{Pols:1998} as presented in \citet{Hurley:2000}. The winds of the \ac{MS} star are given by \citet{Vink:2001}. The \ac{HeMS} winds are parametrised with the multiplicative factor $f_\mathrm{WR}$ as described above and in Appendix \ref{app:winds}. The orbital response to mass loss is given by Eq.~(\ref{eq:wideningWinds}). We go back in the evolution for a \ac{HeMS} lifetime (note that the \ac{HeMS} lifetime depends on how massive the \ac{HeMS} star initially was, which depends on the wind strength we assume) and check that the {\it $M_\mathrm{HeMS} \leq M_\mathrm{HeMS,max}$} condition is satisfied at the start of the HeMS phase and the {\it no RLOF} condition is satisfied throughout this phase.

\begin{figure}
\centering
\includegraphics[width=\columnwidth]{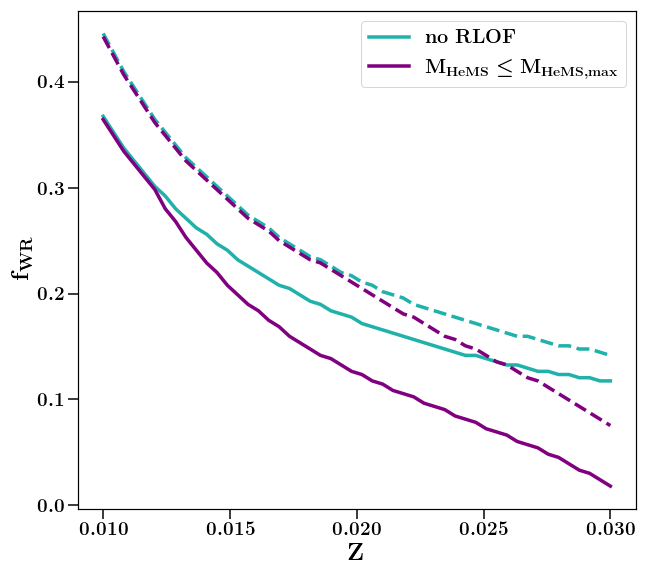}
   \caption{Upper limits on the parametrised Wolf-Rayet wind mass loss rate from \ac{HeMS} stars. The {\it $M_\mathrm{HeMS} \leq M_\mathrm{HeMS,max}$} (purple) and {\it no \ac{RLOF}} (teal) conditions only allow $f_\mathrm{WR}$ values below the curves at metallicity $Z$. The dashed curves correspond to assuming $M_\mathrm{BH}=19.2 M_\odot$ and the current companion Roche-filling factor of $0.990$ rather than the default values of $21.2 M_\odot$ and $0.997$ (solid curves), and indicate the impact of observational uncertainty.}
    \label{fig:constraints}
\end{figure}

Figure \ref{fig:constraints} shows the upper limits on the wind strength, parametrized as $f_\mathrm{WR}$, imposed by these two conditions as a function of metallicity.  Both conditions show that the wind strength has to be reduced from the nominal value $f_\mathrm{WR}=1$ throughout the range of metallicities we have explored. The strongest constraint is placed by the {\it $M_\mathrm{HeMS} \leq M_\mathrm{HeMS,max}$} condition, which is subject to uncertainties in the \ac{MS} wind strengths, overshooting, and the single stellar evolution fits of \citet{Hurley:2000}.  This yields an upper limit $f_\mathrm{WR} \lesssim 0.4$ at $Z=0.01$ and $\lesssim 0.05$ at $Z=0.03$.  However, even if we lift this constraint, the {\it no \ac{RLOF}} condition still places a strong constraint on the allowed mass loss rate: $f_\mathrm{WR} \lesssim 0.45$ at $Z=0.01$ and $\lesssim 0.15$ at $Z=0.03$.  We also explore the impact of \ac{BH} mass and Roche-lobe filling factor measurement uncertainties and find that the constraints on $f_\mathrm{WR}$ change by $\lesssim 0.1$ over the range consistent with observations \citep{MillerJones:2020}.

\section{Cygnus X-1: Future evolution}\label{sec:future}

The revised mass of the companion star in Cygnus X-1 makes it a potential candidate for a future \ac{BH} and raises the intriguing prospect that this system could form a merging binary \ac{BH}, connecting \acp{HMXB} with gravitational-wave sources.  We model the future evolution of the system and find that it is unlikely to form a merging binary \ac{BH}.

\citet{Belczynski:2011} used earlier, lower estimates of the mass of the \ac{BH} and companion in Cygnus X-1 in order to analyse the future evolution of this system.  They predicted that imminent mass transfer from the nearly Roche-lobe filling companion will significantly reduce the companion mass, leaving behind a core that could only form a neutron star, not a \ac{BH}.  They further estimated that the natal kick has a 30\% probability of unbinding the binary, and there was only a $\sim 1\%$ probability that the ensuing neutron star -- black hole binary would merge within 14 Gyr through gravitational-wave emission.

Here we update their predictions based on revised observations, using the COMPAS population synthesis code \citep{Stevenson:2017, VignaGomez:2018} for binary evolution calculations. We estimate the initial masses and orbital separation as in Section \ref{sec:winds}.  We assume a Roche lobe filling factor of 0.997 and a metallicity of $Z=0.02$. The median values for the \ac{BH} mass and the luminosity and temperature of the \ac{MS} companion from \citet{MillerJones:2020} then yield $M_\mathrm{BH}=21.2 M_{\odot}$, $M_\mathrm{opt}=38.9 M_{\odot}$, $a=51.3 R_{\odot}$, and $\tau_\mathrm{MS}=0.81$, i.e., the \ac{MS} companion is 81 per cent through its core hydrogen burning phase by time.  Once the \ac{MS} companion overflows its Roche lobe, the mass transfer phase could brighten Cygnus X-1 up to close to its Eddington luminosity of a few $\times 10^{39}$ erg s$^{-1}$.  Once the envelope is removed, the companion will appear as a \ac{WR} star in an \ac{HMXB} in a phase lasting for $\sim 3 \times10^5$ years.  In the default model described below, we apply \ac{WR} mass loss with $f_\mathrm{WR}=1$ to such a star.

If we follow a treatment similar to \citet{Belczynski:2011}, in which the He core mass is determined by the mass of the star at the end of \ac{MS} (after stripping in this case), we find the companion will collapse into a neutron star.  Even if the binary survives the neutron star natal kick, which happens in 7\% of all binaries in our models, it will generally be too wide to merge through gravitational-wave emission, with only a 0.12\% probability that it will merge in 14 Gyr, similar to the findings of  \citet{Belczynski:2011}.  During this time, it may be detectable in radio pulsar surveys as a neutron-star -- black-hole binary, although the non-recycled pulsar will likely only be observable as a radio source for a few tens of Myr.  However, this treatment may significantly under-predict the core masses of stars that donated mass on the \ac{MS}.

In our revised model, we address the potentially under-estimated core mass of stripped \ac{MS} donors under the assumption that this core mass is determined only by the mass of the star at the end of the \ac{MS}.  We account for the substantial amount of helium synthesized by the companion before interaction with the following crude approximation.  We consider a constant rate of helium production during the \ac{MS}, so the mass of helium produced is $M_\mathrm{He,MS}=\tau_\mathrm{MS} M_\mathrm{c,final}$, where $M_\mathrm{c,final}$ is the final helium core mass that would be achieved at the end of the \ac{MS} phase for the given stellar mass, in the absence of mass transfer. Keeping track of the helium mass synthesized before stripping, $M_\mathrm{He,MS}$, leads to higher core masses at the end of the \ac{MS} phase.  

With this correction, we predict that in the absence of \ac{BH} natal kicks, Cygnus X-1 will become a bound binary \ac{BH}, but one that is too wide to merge within 14 Gyr.  If we incorporate the COMPAS prescription for \ac{BH} kicks based on the ``delayed'' model of \citet{Fryer:2012}, this changes to a 38\% probability of surviving the kick and forming a bound binary \ac{BH}, with a 4\% probability that this binary will merge within 14 Gyr through gravitational-wave emission due to a fortuitous kick. However, the kick prescription for low-mass black holes that do not undergo complete fallback \citep{Fryer:2012} is rather uncertain, with conflicting evidence on the natal kick magnitudes of low-mass \acp{BH} \citep{Repetto:2017,Mandel:2015kicks,Mirabel:2016,Wyrzykowski:2019,Atri:2019}. 

We also consider the impact of varying \ac{WR} mass loss.  Motivated by the results reported in figure \ref{fig:constraints} for $Z=0.02$, we reduce $f_\mathrm{WR}$ to 0.2 from the value of 1 considered above.  This increases the remnant mass of the current optical companion from $\sim 2.9 M_{\odot}$ to $\sim 5.7 M_{\odot}$, and the binary's probability of remaining bound after the supernova from 38\% to 62\%.  With our adjusted prescription for the helium core mass of the stripped companion and $f_\mathrm{WR}=0.2$, Cygnus X-1 has a 5\% probability to merge within 14 Gyr as a binary \ac{BH}.

We thus find that it is possible that a small fraction of \acp{HMXB} like Cygnus X-1 could form merging binary \acp{BH}, although this conclusion is sensitive to the treatment of mass transfer from \ac{MS} donors in population synthesis models and to the natal kick distribution of relatively low-mass \acp{BH}.  If systems like Cygnus X-1 do become progenitors of gravitational-wave events, this would impact the predicted spin distribution of merging binary black holes \citep{Kushnir:2016,Zaldarriaga:2017,Belczynski:2017,FullerMa:2019,Bavera:2019}.  Gravitational-wave observations could ultimately address this possibility by resolving the spin distribution with more events \citep[e.g.,][]{Farr:2017}.  Meanwhile, wide, non-merging binary \acp{BH} could potentially be observable through microlensing \citep{Eilbott:2017}.

\section{Cygnus X-1: Caveats and conundrums}\label{sec:discussion}

We show that the current properties of the Cygnus X-1 system imply a reduction in Wolf-Rayet wind mass loss rates for exposed \ac{HeMS} stars. These results depend on several key assumptions.

We assumed that the optical companion did not experience Roche lobe overflow in its past.  This assumption is consistent with the challenge of detaching from mass transfer once it commences given that the companion has roughly twice the mass of the \ac{BH}, as explained in section \ref{sec:companion}.  However, it is somewhat surprising that several \acp{HMXB} with well measured properties -- Cygnus X-1, LMC X-1 and M33 X-7 -- share not only a high \ac{BH} spin, but also a similar evolutionary state.  Selection effects favour observing bright, long-lived systems, i.e., those with massive main-sequence donors that are close to Roche lobe filling (enabling more efficient accretion).  There may also be an evolutionary stalling point, increasing the number of systems in this phase.  

The latter scenario could indicate that the systems do manage to detach and resume mass transfer multiple times.  While this would negate our wind mass loss rate conclusions, it would imply that much less angular momentum is carried away during non-conservative mass transfer onto a \ac{BH} than we expected.  Assuming non-conservative mass transfer (valid if accretion onto a \ac{BH} is Eddington-limited) from a donor that is twice as massive as the accretor, the specific angular momentum of the material ejected from the binary in units of the binary's specific orbital angular momentum must be $\gamma < 0.85$ in order to avoid a decrease in the size of the Roche lobe (see Eq.~\ref{eq:a_MT} for the change in orbital separation).  For comparison, isotropic re-emission from the \ac{BH} corresponds to $\gamma=2$.   Conversely, if $\gamma=2$, the companion could still disengage from mass transfer if its radius shrinks faster than the size of the Roche lobe in response to mass loss.  This would require the adiabatic logarithmic derivative of radius with respect to mass to exceed $\zeta \equiv d \log R / d \log M > 1.54$, which may be possible for stars in the late phase of their \ac{MS} evolution that have already lost some mass. 

It is also possible that the primary is not fully stripped during the mass transfer episode, but retains about $\sim0.1~M_\odot$ of its hydrogen envelope \citep[e.g.][]{Yoon:2010,Yoon:2017, Bersten:2014,Gotberg:2017,Gotberg:2018,Laplace:2020}. The changed surface abundance could lead to reduced mass-loss rates until the remaining hydrogen is completely removed. However, it is not clear whether retaining an envelope of a fraction of a solar mass could be sufficient to prevent Wolf-Rayet-like winds. In any case, whether Wolf-Rayet wind mass loss rates must be lower than anticipated or whether stars that experience mass transfer in binaries are only partially stripped, the impact on binary evolution is similar: there is less mass loss than previously assumed.  In fact, our Wolf-Rayet wind reduction factors can be broadly interpreted as constraints on winds from stripped stars, whether they are naked helium stars or retain a small hydrogen-rich envelope.

Naked helium cores can expand significantly in the last stages of their lives, potentially leading to another mass transfer episode from the \ac{BH} progenitor late in the evolution. However, the degree of expansion is very mild for stars with initial masses $\gtrsim 20 M_\odot$ at near-solar metallicities \citep{Yoon:2010,Hirai:2017,Laplace:2020}, so Cygnus X-1 is unlikely to have experienced such mass transfer.

As a consequence of its mass accretion history, the secondary may be over-luminous relative to  single stars of the same total mass \citep[][but see \citealt{Hellings:1983}, who concludes that they \ac{MS} accretors quickly return to single-star models, and \citealt{BraunLanger:1995}, who reach the opposite conclusion and find that accretors are under-luminous]{DrayTout:2007}.  Since we use single star evolutionary tracks to estimate the properties of the secondary star, this can affect our wind constraint from the no-RLOF condition.  However, since we choose a stellar model that matches the observed radius of the secondary at the present day, we do not anticipate the impact to be significant.

Finally, as figure \ref{fig:constraints} shows, the level of reduction in the winds is sensitive to the assumed metallicity of Cygnus X-1.  We now discuss this in more detail.

\citet{Shimanskii:2012} report that helium, carbon, oxygen, aluminium, sulfur and iron are overabundant  by [X/H]= 0.23--0.43 dex compared to the solar values \citep[]{Anders:1989}. Nitrogen, neon, and silicon have an even higher overabundance of [X/H]=0.69--0.94 dex. These values appear robust against variations due to orbital motion and Roche-lobe filling factors, although some hydrogen and helium lines are sensitive to variations in the wind \citep{Shimanskii:2012}.  

Previous accretion from the \ac{BH} progenitor could significantly alter the chemical profile on the surface of the companion.  For example, the detailed models of \citet{Qin:2019} predict the observed enhancement of companion nitrogen abundances as a consequence of late main sequence mass transfer from the \ac{BH} progenitor.  In addition to direct accretion, which is expected to enhance helium and nitrogen abundances, the deposited angular momentum can lead to a dramatic spin-up of the \ac{MS} star \citep[e.g.][]{Packet:1981}, although spin-up to near break-up frequencies may suppress subsequent accretion. The surface could then be enhanced by helium and CNO-elements due to rotational mixing \citep{Meynet:2000, Heger:2000, Przybilla:2010}. The rotational mixing might also make the star over-luminous compared to a non-rotating model \citep{Langer:1992}.  The optical companion is observed to be tidally locked at present, with an inferred ratio of the rotational to orbital frequency of $1.05 \pm 0.10$ \citet{MillerJones:2020}.  Assuming a present-day rotational period of 5.6 days, the rotational frequency is a third of the Keplerian (break-up) frequency at the stellar equator).  Alternatively, as discussed above and contrary to the channel assumed in this work, the \ac{MS} companion may have been partially stripped by mass transfer onto the \ac{BH} or its progenitor after the initial mass transfer phase from the primary, revealing deeper layers of the star.

Although these mechanisms could be responsible for the overabundance of some of the elements, they have difficulty in explaining the overabundance of late stage burning elements such as silicon and iron. This implies that these elements were primordially enhanced or were deposited from the progenitor of the \ac{BH} in the final stages of its evolution. 

High primordial abundances imply that both stars in the binary had a high metallicity at birth, which therefore requires a very strong reduction in the mass-loss rate (a factor of $\sim 10$) following the constraints described in section \ref{sec:winds}.  On the other hand, a weak explosion induced by the collapse of the core can lead to an ejection of a small fraction of the outer part of the envelope at very low velocities. Because most of the envelope is assumed to fall back into the \ac{BH}, the ejected material will be barely above the escape velocity, and could be efficiently accreted by the companion in a \ac{RLOF}-like manner. If the heavy elements synthesized at the centre are efficiently mixed up to the outer regions before the inner slower material starts falling back, these elements can accrete onto the surface of the secondary. Such efficient mixing of heavy elements has been observed in supernovae such as SN1987A and Cassiopeia A \citep[e.g.][]{Utrobin:1995,Fesen:2006}, and has been reproduced in 3D supernova explosion simulations \citep[e.g.][]{Hammer:2010,Wongwathanarat:2015,Wongwathanarat:2017}, while \citet{Liu:2015,Hirai:2018} explore the contamination of a \ac{MS} companion by supernova ejecta. However, it is not clear whether similar degrees of mixing can be induced in failed supernovae that form \acp{BH} rather than neutron stars, and the abundance pattern of the Cygnus X-1 companion merits further investigation.

Regardless of whether we assume that the observed companion metallicity of $Z=0.03$ \citep{Shimanskii:2012} is primordial, or use the lower metallicity of $Z=0.01$ based on HD~227460 \citep{Daflon:2001}, we conclude that the observed properties of Cygnus X-1 require a reduction in Wolf-Rayet winds to $\sim 5$--40\% of their previously assumed values in the context of our assumed evolutionary channel. 

Recent theoretical modelling of mass-loss from stripped stars \citep{Vink:2017,Sander:2020,SanderVink:2020} points to reduced mass-loss rates compared to earlier literature. Moreover, these models suggest a steep dependence on the Eddington factor, which can change significantly during the lifetime of the stripped star. This indicates that extrapolating the empirical Wolf-Rayet mass-loss rates to the entire  duration of the stripped star life is misleading. Our results are qualitatively consistent with these findings. The reduced mass-loss rates could also be attributed to strong wind clumping, which is expected to occur in line-driven winds due to radiative instabilities \citep[]{Owocki:1988,Sundqvist:2018}. Clumping of winds has been indirectly observed for massive MS stars in X-ray binaries \citep[]{ElMellah2018,Lomaeva:2020} and stripped stars may also experience high degrees of clumping.

We further find that \acp{HMXB} like Cygnus X-1 form through a different evolutionary channel than the bulk of merging binary black holes \citep[see, e.g.,][for a review]{MandelFarmer:2018}.  However, a fortuitous natal kick accompanying the birth of the secondary \ac{BH} could lead Cygnus X-1 to merge as a \ac{BH} binary within 14 Gyr.  Gravitational-wave observations may be able to constrain the contribution of this channel to the formation of merging binary \acp{BH} through spin measurements.

\section*{Acknowledgements}
Simulations in this paper made use of the COMPAS rapid binary population synthesis code, which is freely available at \url{http://github.com/TeamCOMPAS/COMPAS}.  We thank Alexander Heger, Hagai Perets, Jerry Orosz, Stephen Justham, Selma de Mink and Jakub Klencki for discussions, and Robert Izzard and Silvia Toonen for comments on a preliminary draft of the manuscript. C.~J.~N.~thanks the University of Birmingham for financial support. JCAM-J and IM are recipients of Australian Research Council Future Fellowships (FT140101082 and FT190100574, respectively).
AVG acknowledges funding support by the Danish National Research Foundation (DNRF132).




\bibliographystyle{mnras}
\bibliography{bibliography} 

\newpage


\appendix
\section{Wind mass-loss rate for stripped stars}\label{app:winds}

In this appendix, we describe the particular parametrized formalism we used to model wind-driven mass loss from stripped stars.  While more recent theoretical models are available \citep[e.g.,][]{Vink:2017,Sander:2020}, this provides us with a convenient framework for investigating the empirical constraints placed by Cygnus X-1.

\citet{Hamann:1995} formulated a prescription for the wind mass loss rate for stripped or Wolf-Rayet stars as a function of the mass and luminosity of the stripped star. \citet{Hamann:1998} expanded on this work by reducing winds by a factor of $\sqrt{D}$ where $D$ is the wind clumping factor \citep{Moffat:1988,Nugis:1998}.   \citet{Vink:2005} further introduced a metallicity dependence to the winds.  Combining the effects of clumping and metallicity leads to the prescription
\begin{equation}\label{eq:dMdtWR}
(dM/dt)_\mathrm{WR} ={\underbrace{\frac{1}{\sqrt{D}}10^{-11.95} \frac{L}{L_{\odot}}^{1.5}}_1}{\underbrace{ \frac{Z}{Z_{\odot}}^{0.86}}_2}\ \mathrm{M_{\odot}\ yr^{-1}},
\end{equation}
where term 1 is the result of \citet{Hamann:1995, Hamann:1998} and term 2 is from \citet{Vink:2005}.
Setting $D=100$, i.e., reducing winds by a factor of 10, recovers Eq.(9) of \citet{Belczynski:2010} and is consistent with the winds of \citet{Yoon:2010}.

We follow \citet{Barrett:2017FIM} in scaling the prescription of \citet{Belczynski:2010} by a multiplicative factor $f_\mathrm{WR}$ in order to parametrise the uncertainty in the wind mass-loss rates:
\begin{equation}\label{eq:wr}
(dM/dt)_\mathrm{WR} = f_\mathrm{WR}\times 10^{-13} \frac{L}{L_{\odot}}^{1.5} \frac{Z}{Z_{\odot}}^{0.86}\ \mathrm{M_{\odot}\ yr^{-1}}.
\end{equation}
Note that this is not the same $f_\mathrm{WR}$ as used in \citet{Yoon:2010} because ours already assumes a reduction of the original wind prescription of \citet{Hamann:1995}.
The default assumption of $f_\mathrm{WR}=1$ corresponds to the default models of \citet{Belczynski:2010, Yoon:2010, Stevenson:2017, Qin:2019}.

It is challenging to interpret our constraints on $f_\mathrm{WR}$ in terms of a wind clumping factor $D$. A direct interpretation of $f_\mathrm{WR}=0.2$ would imply a clumping factor of $D=2500$ in the model of Eq.~(\ref{eq:dMdtWR}). \citet{Nugis:1998, Nugis:2000} used a clumping factor ranging from 10 to 30; $D$ could be as high as 16 according to \citet{Hamann:1998}. 
More recent theoretical models by \citet{Sander:2017} use depth-dependent clumping factors and suggest a maximum value at infinity of $D_\infty=10$ to be consistent with observations of electron scattering wings.  In later works however,  \citet{Sander:2020} and \citet{SanderVink:2020} propose 50 as a maximum upper limit for $D_\infty$, by comparison with previous theoretical works \citep{Grafener:2005} and O/B star analyses \citep[e.g.][]{Bouret:2012,Mahy:2015}. 
A clumping factor of $2500$ seems extra-ordinary compared to previous models. 
As mentioned in the Section \ref{sec:discussion}, additional constraints on the clumpiness of stellar winds from massive stars can be obtained from X-ray binaries \citep[e.g.][]{Lomaeva:2020,ElMellah2018, Grinberg:2017}.

In light of the above, our reduction of $f_\mathrm{WR}$ is probably best interpreted as an overall constraint on mass loss from massive stripped stars \citep{Vink:2017}, perhaps indicating a different dependence on metallicity or luminosity \citep[cf.][]{Sander:2020,SanderVink:2020}, rather than a specific change in the clumping.

\bibliographystyle{aasjournal}
\end{document}